\documentclass[aps,pra,twocolumn,amsmath,amssymb,nofootinbib,showpacs,superscriptaddress]{revtex4-1}
\usepackage[english]{babel}
\usepackage{latexsym}
\usepackage{graphics}
\usepackage{graphicx}
\usepackage{epsfig}
\usepackage{color}
\usepackage{bm}
\usepackage{amsmath}
\usepackage{amssymb}
\usepackage{amsthm}
\usepackage{dcolumn}
\usepackage{bm}
\usepackage{float}
\usepackage{hyperref}
\usepackage{color}
\usepackage{epstopdf}
\usepackage{braket}
\usepackage{cleveref}
\usepackage[svgnames]{xcolor}
\hypersetup{hidelinks,colorlinks=true,allcolors=DarkBlue}

\newcommand{\CZ}{\textsf{C}$Z$}
\newcommand{\CX}{\textsf{C}$X$}
\newcommand{\CZth}{\textsf{C}$Z_{\theta}$}
\newcommand{\CnZ}{\textsf{C}$^{N-1}Z$}
\newcommand{\CnX}{\textsf{C}$^{N-1}X$}
\newcommand{\CnU}{\textsf{C}$^{N-1}U$}
\newcommand{\CnZth}{\textsf{C}$^{N-1}Z_{\theta}$}
\newcommand{\CnUb}{\textsf{C}$^{N}{\bf U}$}
\newcommand{\CUb}{\textsf{C}${\bf U}$}

\theoremstyle{remark}

\begin{document}

\preprint{APS/123-QED}

\title{Scalable quantum computing with qudits on a graph}

\author{E.O. Kiktenko}
\affiliation{Russian Quantum Center, Skolkovo, Moscow 143025, Russia}
\affiliation{Steklov Mathematical Institute of Russian Academy of Sciences, Moscow 119991, Russia}
\affiliation{Moscow Institute of Physics and Technology, Dolgoprudny, Moscow Region 141700, Russia} 

\author{A.S. Nikolaeva} 
\affiliation{Russian Quantum Center, Skolkovo, Moscow 143025, Russia}
\affiliation{Moscow Institute of Physics and Technology, Dolgoprudny, Moscow Region 141700, Russia} 

\author{Peng Xu}
\affiliation{State Key Laboratory of Magnetic Resonance and Atomic and Molecular Physics,
Wuhan Institute of Physics and Mathematics, Chinese Academy of Sciences --- Wuhan National Laboratory for Optoelectronics, Wuhan 430071, China}
\affiliation{Center for Cold Atom Physics, Chinese Academy of Sciences, Wuhan 430071, China}

\author{G.V. Shlyapnikov}
\affiliation{Russian Quantum Center, Skolkovo, Moscow 143025, Russia}
\affiliation{Moscow Institute of Physics and Technology, Dolgoprudny, Moscow Region 141700, Russia} 
\affiliation{LPTMS, CNRS, Univ. Paris-Sud, Universit\'e Paris-Saclay, Orsay 91405, France}
\affiliation{SPEC, CEA \& CNRS, Universit\'e Paris-Saclay, CEA Saclay, Gif-sur-Yvette 91191, France}
\affiliation{Van der Waals-Zeeman Institute, Institute of Physics, University of Amsterdam, Science Park 904, 1098 XH Amsterdam, The Netherlands}

\author{A.K. Fedorov}
\affiliation{Russian Quantum Center, Skolkovo, Moscow 143025, Russia} 
\affiliation{Moscow Institute of Physics and Technology, Dolgoprudny, Moscow Region 141700, Russia} 

\date{\today}
\begin{abstract}
We show a significant reduction of the number of quantum operations and the improvement of the circuit depth for the realization of the Toffoli gate by using qudits. 
This is done by establishing a general relation between the dimensionality of qudits and their topology of connections for a scalable multi-qudit processor, where higher qudit levels are used for substituting ancillas.
The suggested model is of importance for the realization of quantum algorithms and as a method of quantum error correction codes for single-qubit operations. 
\end{abstract}

\maketitle

\section{Introduction}

Remarkable progress in realizing controllable quantum systems of an intermediate scale~\cite{Lukin2017,Monroe2017,Browaeys2018,Martinis2018,Blatt2018} 
makes it realistic to study properties of strongly correlated quantum matter~\cite{Trotzky2012,Mazurenko2017,Lukin2019,Blatt2019} 
and to implement various quantum algorithms~\cite{Montanaro2016,Martinis2016,Gambetta2017,Blatt20182,Gambetta2019}.
However, existing quantum computing systems lack either coherence or controllable interactions between qubits, and this limits their capabilities. 
A serious obstacle in realizing quantum algorithms is a large number of two-qubit gates, which requires programmable inter-qubit interactions and can cause decoherence.
The situation becomes even more challenging in the case of mulit-qubit gates, such as an $N$-qubit Toffoli gate, 
which is a basic building block for quantum algorithms like Shor's algorithm~\cite{Shor1997} and for quantum error corrections schemes~\cite{Shor1996,Cory1998,Reed2012}.
Its implementation requires $12N-23$ two-qubit gates with $N-2$ ancilla qubits or $\mathcal{O}(N^2)$ gates without them~\cite{Barenco1995}, 
which is of high cost for near-term noisy intermediate-scale quantum devices. 
Therefore, the reduction of the number of operations that are required for the realization of multi-qubit gates remains a crucial problem.

One of the possible ways to reduce the number of required operations is to use additional degrees of freedom of quantum systems.
This idea has stimulated an extended activity~\cite{Ruben2018,Zeilinger2018} in 
theoretical~\cite{Farhi1998,Muthukrishnan2000,Nielsen2002,Berry2002,Klimov2003,Bagan2003,Vlasov2003,Clark2004,Ralph2007,Ivanov2012,Kiktenko2015,Kiktenko20152,Pavlidis2017,Bocharov2107,Gokhale2019,Ionicioiu2019,Senko2019} 
and experimental studies~\cite{Martinis2009,White2009,Wallraff2012,Gustavsson2015,Katz2015,Ustinov2015,Morandotti2017,Balestro2017} of quantum computing models with qudits, which are $d$-dimensional ($d>2$) quantum systems. 
In particular, qudits can be used for substituting ancillas~\cite{Ralph2007,White2009,Gokhale2019,Ionicioiu2019}, which allows the reduction of the required number of interactions between information carriers for the realization of multi-qubit gates. 
In experiments with photonic quantum circuits~\cite{White2009}, for a system of an $N$-dimensional qudit connected with $N-1$ qubits, the $N$-qubit Toffoli gate was realized with $2N-3$ qubit-qudit gates.
However, it is hard to expect scalability for such  a system with increasing $N$, although qudits with $d$ up to 10 have been realized~\cite{Morandotti2017}.
Alternative schemes allow further reduction in the number of operations~\cite{Ionicioiu2019} or circuit depth~\cite{Gokhale2019}. 
However, they require either additional measurement-based feedforward corrections or specific topology with (almost) all-to-all connectivity.
It should be noted that qudits can be also used for optimizing the resources in quantum communications~\cite{Gisin2002,Brus2002,Kaszlikowski2003}.

\begin{figure}
	\includegraphics[width=\linewidth]{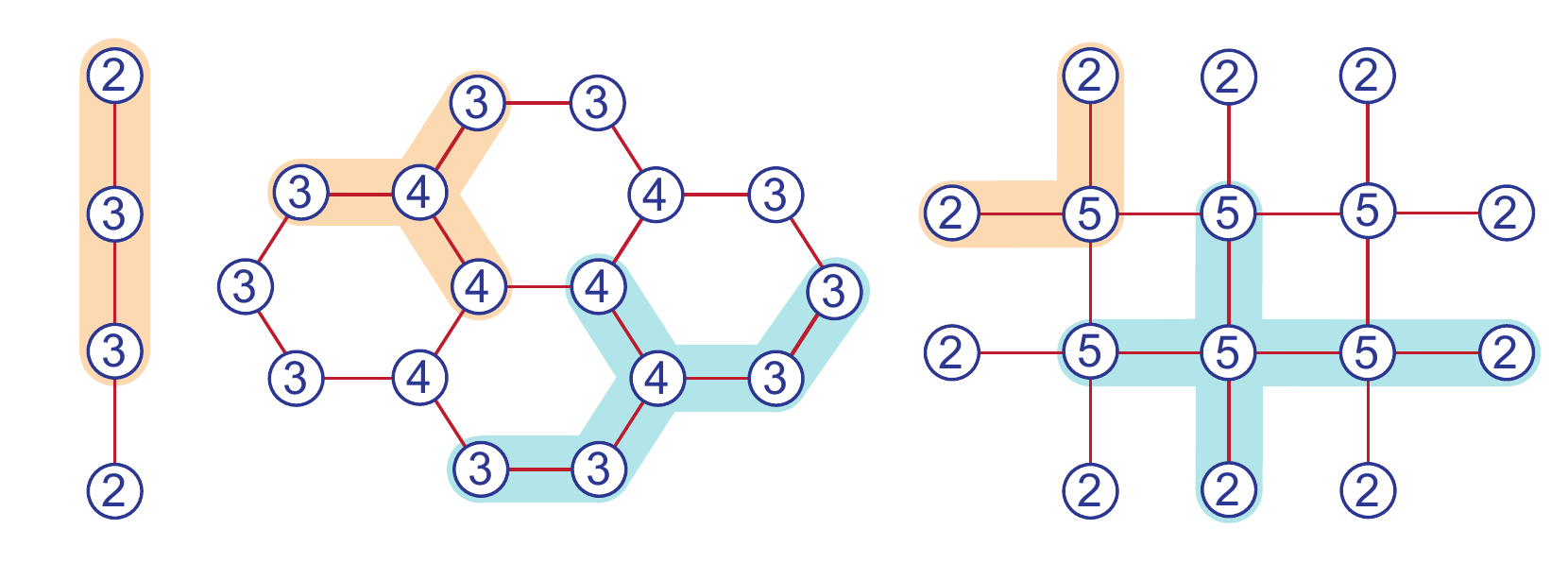}
	\vskip-4mm
	\caption{Illustration of various multi-qudit and qubit-qudit schemes for quantum computing that provides reduction in the number of operations for the realization of the Toffoli gate and reduces the depth of the corresponding circuit.
	We note that if condition~\eqref{eq:main_condition} holds for a whole system, then it is also fulfilled for {\it any} of its acyclic-connected subsystem (highlighted).}
	\label{fig:opt_topology}
\end{figure}

In this work, we study a scalable quantum computing model based on qudits, which uses higher qudit levels as ancillas.
For this model we establish a general relation between the dimensionality of qudits and the topology of qudit connectivity: 
for a given qudit one should have $d\geq k+1$, where $k$ is the number of links of this qudit with the others.
We then demonstrate that this is the key relation for achieving the best-known performance in the number of operations without additional measurement-based corrections.
The obtained results are useful for ongoing experiments with quantum computing systems of various nature, such as 
Rydberg atom arrays~\cite{Lukin2017,Browaeys2018}, trapped ions~\cite{Monroe2017,Blatt2018,Blatt20182,Blatt2019}, 
integrated optics~\cite{White2009,Morandotti2017}, and superconducting circuits~\cite{Martinis2018,Martinis2009,Wallraff2012,Gustavsson2015,Katz2015,Ustinov2015}. 

The paper is organized as follows. 
In Sec.~\ref{sec:qudits}, we describe a model of qudit-based processor and formulate the necessary condition for efficient implementation of the $N$-qubit Toffoli gate.
In Sec.~\ref{sec:Toffoli}, we consider a circuit construction for the implementation of the $N$-qubit Toffoli gate.
In Sec.~\ref{sec:multiqudit}, we generalize our results for multi-qubit controlled unitary gates.
In Sec.~\ref{sec:experiment}, we briefly discuss possible experimental realization of our scheme.
We summarize our results and provide outlook in Sec.~\ref{sec:concl}.

\begin{figure*}
	\includegraphics[width=1.0\linewidth]{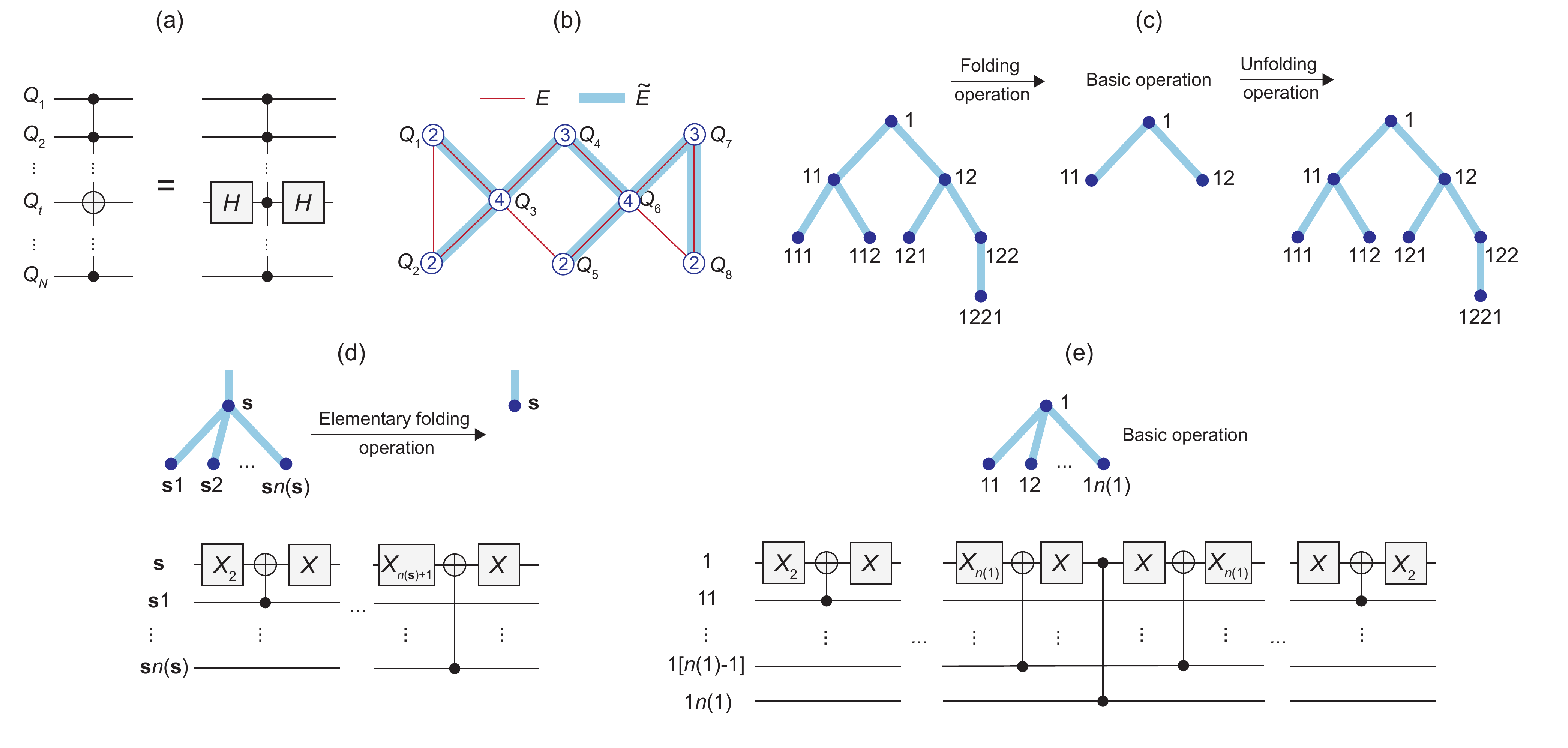}
	\vskip-4mm
	\caption{
		In (a) we present the decomposition of $N$-qubit Toffoli gate \CnX$^{(t)}$ in \CnZ~and Hadamard gates.
		In (b) an example of the connection topology of an eight-qudit system is shown. 
		Here $\widetilde{E}\subset E$ is a subset of connections, which correspond to the connected acyclic graph.
		Numbers in the nodes correspond to minimal dimensions of qudits determined by the general condition~\eqref{eq:main_condition}.
		In (c) the action of three general steps in the tree representation is presented.
		The first step is the folding operation, which wraps the original tree to the single-level form. 
		The second step is the basic operation, which does not change the tree.
		The third step is the unfolding operation that returns the tree to its original form.
		In (d) the elementary folding operation with the corresponding circuit is shown.	
		In (e) the circuit corresponding to the basic operation for the \CnZ~gate is presented.
	}\label{fig:main}
\end{figure*}

\section{Qudit processor: Optimal relation between dimensionality and topology} \label{sec:qudits}

Consider a system of $N$ qudits denoted as $Q_i$, $i\in\{1,\ldots,N\}$.
Let each qudit $Q_i$ have dimensionality $d_i \geq 2$.
In our setup we consider the first two levels $\ket{0}_{Q_i}$ and $\ket{1}_{Q_i}$ as qubit levels, and higher levels $\ket{n}_{Q_i}$ with $n\geq 2$ as auxiliary levels. 

We assume that the initial state of our $N$-qudit system can be considered as an $N$-qubit state, i.e. the system $Q_1\ldots Q_N$ is in a pure or mixed state with zero population of auxiliary levels for each of the qudits $Q_i$.
We then determine the set of operations that can be performed within the system. 
In analogy with the idea of qubit-based universal quantum computations, we assume that we are able to perform any desirable unitary operation on a two-level subspace spanned by the qubit level states $\ket{0}_{Q_i}$ and $\ket{1}_{Q_i}$.
Meanwhile, these single-qubit operations act as identity operators in the spaces of auxiliary levels. 
At the same time, we assume an ability to perform two-qubit \CZ~gates corresponding to certain topology of physical connections between qudits.
To determine this topology we introduce a set $E$ of ordered pairs $(i,j)$, such that $i,j\in\{1,\ldots,N\}, i<j$. 
We assume that if $(i,j)\in E$, then one is able to perform the operation,
\begin{equation}\label{eq:CZ_gate_nearest}
\begin{aligned}
&\textsf{C}Z\ket{11}_{Q_i,Q_j} = -\ket{11}_{Q_i,Q_j}\\
&\textsf{C}Z\ket{xy}_{Q_i,Q_j} = \ket{xy}_{Q_i,Q_j} \text{ for } xy\neq 1,
\end{aligned}
\end{equation}
with $x\in\{0,\ldots, d_i-1\}$ and $y\in\{0,\ldots, d_j-1\}$.
We also assume that $E$ corresponds to the $N$-vertex-connected graph, i.e. there is a path between any pair of qudits.

We note that the \CZ~gate can be easily transformed to the more common controlled-not \CX~gate using two Hadamard gates.
Finally, we consider the ability to manipulate the auxiliary levels.
We assume that one is able to apply a generalized inverting gate
\begin{equation}
	\begin{aligned}
		& X_m\ket{0}_{Q_i} = \ket{m}_{Q_i}, \quad X_m\ket{m}_{Q_i} = \ket{0}_{Q_i}, \\
		&X_m\ket{y}_{Q_i} =\ket{y}_{Q_i} \text{ for } y\neq 0,m
	\end{aligned}
\end{equation}
to every qudit $Q_i$.
We note that $X_1$ is actually the standard qubit $X$ gate, and $X_m$ is the only operation engaging auxiliary qudit levels in our setup.

Let $\widetilde{E}\subseteq E$ define an $N$-vertex connected acyclic graph known as a tree.
We note that $\widetilde{E}$ can be always obtained from $E$ by eliminating connections in the case of cycles in the original graph defined by $E$. 
This can be done, e.g. by keeping all edges explored by the depth-first search (DFS) algorithm~\cite{Kleinberg2006} and removing unexplored ones during $E$ traversal. 
We note that the complexity of the DFS algorithm is known to be $\mathcal{O}(L)$, where $L=N+\frac{1}{2}\sum_{i=1}^{N}k_i$ is the total number of nodes and edges of the graph.

Our main result is the demonstration that a strong reduction in the number of operations required for the realization of the $N$-qubit Toffoli gate is possible if the following relation between the dimensionality of a qudit $d_i$ and the number $k_i$ of its connections to other qudits within $\widetilde{E}$ is satisfied: 
\begin{equation} \label{eq:main_condition}
	d_i \geq k_i+1.
\end{equation}
In what follows we show that if this condition is fulfilled, then it is possible to realize the $N$-qubit Toffoli gate by employing $2N-3$ two-qudit \CZ~gates~\eqref{eq:CZ_gate_nearest}. 

This result gives a general picture of simplifying quantum logical operations on qudit-based processors.
Let condition~\eqref{eq:main_condition} be satisfied with $k_i$ replaced with a number of connections of $Q_i$ to other qudits within the full connection set $E$ instead of acyclic subset $\widetilde{E}$.
In this case, condition~\eqref{eq:main_condition} is automatically satisfied for any connected acyclic subgraph consisting of $M<N$ nodes.
This means that the $M$-qubit Toffoli gate can be efficiently implemented for any connected subset of $M$ qudits from $\{Q_i\}$.
Then the condition~\eqref{eq:main_condition} opens a way to formulate a desirable relation between the dimension of employed qudits and topology of their connections.
Let us illustrate this relation for specific cases: it is preferable to employ qutrits ($d_i=3$)  for the linear topology, ququarts ($d_i=4$) for honeycomb topology, qukwints ($d_i=5$) for a 2D rectangular lattice, and so on (see Fig.~\ref{fig:opt_topology}).

\section{Toffoli gate implementation} \label{sec:Toffoli}

The generalized $N$-qubit Toffoli gate \CnX$^{(t)}$ flips a particular target qubit state of $Q_t$ if and only if all other $N-1$ control qubits are in the state 1.
The main operation behind \CnX$^{(t)}$~gate is the following \CnZ~operation:
\begin{equation}\label{eq:global_CZ_gate}
\begin{aligned}
	&\textsf{C}^{N-1}Z\ket{1\ldots 1}_{Q_1\ldots Q_N} = -\ket{1\ldots 1}_{Q_1\ldots Q_N},\\
	&\textsf{C}^{N-1}Z\ket{x_1\ldots x_N}_{Q_1\ldots Q_N} = \ket{x_1\ldots x_N}_{Q_1\ldots Q_N},
\end{aligned}
\end{equation}
for $\prod_ix_i\neq 1$ [see Fig.~\ref{fig:main}(a)].
This operation does not depend on $t$.
The choice of the target qubit $t$ can be maid by adding single-qubit Hadamard gates.

The generalized $N$-qubit Toffoli gate costs $2N-3$ two-qudit \CZ~gates.
To demonstrate this fact we use the $N$-vertex acyclic graph $\widetilde{E} \subset E$ [see Fig.~\ref{fig:main}(b)], which satisfies condition~\eqref{eq:main_condition}.
We note that in the particular example in Fig.~\ref{fig:main}(b) condition~\eqref{eq:main_condition} is satisfied for $\widetilde{E}$ but not for $E$.

We start with representing $\widetilde{E}$ as a tree that is always possible for any acyclic connected graph.
As we show below, the optimal node to choose as a root in the tree representation is a node that provides the minimal height of the resulting tree, i.e. in this case the number of edges between the root and the farthest node is minimal. 
In order to find an optimal node for the root one can consistently apply a leaves-reduction operation that removes nodes of unit degree (nodes with only one edge) from an input graph.
After a number of such operations, a graph consisting of either one or two connected nodes is obtained.
Each of the nodes of the final graph can be chosen as an optimal root for the original graph.
The complexity of this algorithm is $\mathcal{O}(N)$ since each node is accessed only once.

We use the following rules for tree node notations.
We mark each node with a string consisting of integer numbers: the root is denoted with 1; 
the siblings of node ${\bf s}$ are denoted as ${\bf s}1, {\bf s}2, \ldots {\bf s}n({\bf s})$, where $n({\bf s})$ is the total number of node ${\bf s}$ siblings 
[see an example in Fig.~\ref{fig:main}(c)].

The realization of the \CnZ~operation is related to operations with the tree and consists of three main steps: (i) folding operation, (ii) basic operation, and (iii) unfolding operation [see Fig.~\ref{fig:main}(c)].
First, we realize the folding operation in order to bring the original tree into a single-level form, where the root siblings do not have any siblings themselves. 
This is achieved by applying the sequence of elementary folding operations [see Fig.~\ref{fig:main}(d)].
Together with each of these operations, we perform a sequence of gates on qudits corresponding to the nodes involved in this particular elementary folding operation.
The sequence of gates is depicted in the bottom part of Fig.~\ref{fig:main}(d).
For each of the leaves ${\bf s}i$ ($i\in\{1,\ldots,n({\bf s})\}$), we implement the following sequence of three gates: 
(i) the $X_{1+i}$ gate of the parent node qudit ${\bf s}$; 
(ii) the \CX~gate with ${\bf s}i$ being a control and ${\bf s}$ being a target; 
(iii) the additional $X$ gate on ${\bf s}$.
This sequence of gates leaves qudit ${\bf s}$ in the state $\ket{1}_{\bf s}$ if and only if ${\bf s}$ and ${\bf s}i$ initially were in the state $\ket{11}_{{\bf s},{\bf s}i}$.
Finally, the elementary folding operation on a subtree ${\bf s}$, ${\bf s}1,\ldots,{\bf s}n({\bf s})$ keeps the qudit ${\bf s}$ in the state $\ket{1}_{\bf s}$ 
if and only if all qudits ${\bf s}$, ${\bf s}1,\ldots, {\bf s}n({\bf s})$ are in the state 1 before its start.
Otherwise, the qudit ${\bf s}$ turns into the state 0 or into a state related to auxiliary levels.
We note that the elementary folding operation preserves computational basis states, 
and requirement~\eqref{eq:main_condition} guarantees that the number of additional levels is sufficient to perform all required $X_m$ operations.
By considering the evolution of the arbitrary $N$-qubit computational basis state during the whole folding operation we obtain that the root siblings $11,\ldots,1n(1)$ turn into the state $\ket{1\ldots1}_{11,\ldots,1n(1)}$ 
if and only if all the qudits except the root are initialized in the state 1.

\begin{figure*}
	\includegraphics[width=0.9\linewidth]{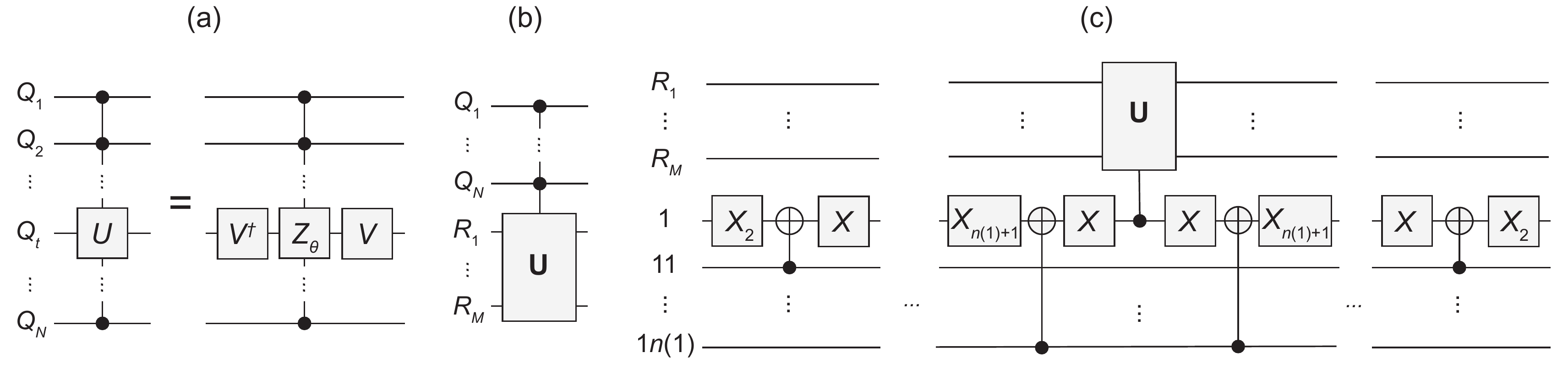}
	\caption{
		In (a) the decomposition of \CnU$^{(k)}$ operation using the spectral decomposition of $U$ is shown.
		In (b) the \CnUb$^{(R_1,\ldots,R_M)}$~gate is shown.
		In (c) the circuit corresponding to the basic operation for the \CnUb~gate is shown.
	}\label{fig:generalization}
\end{figure*}

At the second step, we implement an operation that only performs a sequence of gates on qudits, which correspond to the root and its leaves [Fig.~\ref{fig:main}(e)].
We note that this operation does not modify the tree structure.
This is achieved by implementing sequences of gates similar to the elementary folding operations on the leaves $11, \ldots, 1[n(1)-1]$ and applying the \CZ~gate to the root and to the last leaf $1n(1)$.
Finally, we perform the previous sequence of gates in the reverse order.
The resulting sequence of gates then corresponds to the following transformation of the computational basis states:
\begin{equation}
\begin{aligned}
	&\ket{11\ldots1}_{1,11,\ldots,1n(1)}\rightarrow -\ket{11\ldots1}_{1,11,\ldots,1n(1)}\\
	&\ket{yx_1\ldots x_{n(1)}}_{1,11,\ldots,1n(1)}\rightarrow \ket{yx_1\ldots x_{n(1)}}_{1,11,\ldots,1n(1)}
\end{aligned}
\end{equation}
for $y\prod_{i=1}^{n(1)}x_i\neq1$.
In other words, the computational basis state of the whole $N$-qudit state after the folding operation accumulates an additional phase factor $-1$, if all the qudits after the basic operation were in the state 1, and remains unchanged otherwise. 

Eventually, we perform the unfolding operation, which is the folding operation in the reverse order.
It transfers computational basis states after the folding operation back to their initial form.
We note that as a result we obtain an $N$-qubit state.
However, due to the basic operation, the state $\ket{1\ldots1}_{Q_1,\ldots, Q_N}$ accumulates the additional phase factor $-1$ after all three steps.
This is exactly the desired operation \eqref{eq:global_CZ_gate}.

One can see that the number of employed two-qubit \CZ~gates is $2N-3$. 
Each of the qudits corresponding to tree nodes, except for the root and the root last sibling $1n(1)$, serves as a  control qubit in \CX~operations twice (in the folding and unfolding steps). 
Besides that, there is a single \CZ~operation between the root and $1n(1)$ node qudit.

We note that elementary (un)folding operations can be performed in parallel, whereas realizing quantum gates inside these operations cannot be parallelized.
Thereby the depth of the resulting circuit is determined by the height of the tree and the number of elements in each of its levels.
Thus, it is preferable to choose the root such that the height of the tree is minimal.
We also note that the depth of the resulting circuit is highly dependent on the particular topology of the underlying tree.
This fact makes it difficult to compare directly a circuit depth resulting from our approach with the one proposed in Ref.~\cite{White2009}, where the same number of nonlocal gates $2N-3$ is employed.
However, we can conclude that the depth of the circuit constructed for a complete $\kappa$-ary tree, with a fixed parameter $\kappa$, belongs to $\mathcal{O}(\log N)$.
It is achieved by parallelizing (un)folding operations for $\kappa$ subtrees for all levels except the top.
In contrast, the depth of the circuit constructed according to the approach of Ref.~\cite{White2009}, which considers a root connected to $N-1$ leaves, belongs to $\mathcal{O}(N)$ since no parallelizing can be applied.

\section{Multi-qudit generalization} \label{sec:multiqudit}

Our approach can be further generalized for the implementation of multi-qubit controlled unitary gate \CnU$^{(t)}$, 
where the target qubit state $Q_t$ goes through a single-qubit unitary operation $U$ if all control qubits are in the state 1 [see Fig.~\ref{fig:generalization}(a)].
It can be realized using a spectral decomposition of $U$ in the form $U=VZ_{\theta}V^\dagger$, where $V$ is a certain unitary operator and $Z_{\theta}\equiv \ket{0}\bra{0}+e^{{\rm i}\theta}\ket{1}\bra{1}$ for some value of $\theta$.
Then the implementation of \CnU$^{(k)}$~reduces to the implementation of \CnZth~and single qubit $V$ and $V^{\dagger}$ operations [see Fig.~\ref{fig:generalization}(a)].
\CnZth~can be implemented in the same way as \CnZ~with the only difference being that in the basic operation given in Fig.~\ref{fig:main}(e) the central \CZ~gate has to be replaced with the \CZth~gate.
Depending on the concrete physical realization of the computing platform, \CZth~can be performed either directly or being decomposed into two \CZ~gates and local operations~\cite{Barenco1995}.
Thus, one needs either $(2N-4)$ \CZ~gates plus a single \CZth~gate or $(2N-2)$ \CZ~gates.

We also consider a realization of \CnUb$^{(R_1,\ldots,R_M)}$~operation [Fig.~\ref{fig:generalization}(b)], which performs an $M$-qubit unitary operator ${\bf U}$ on qubits (or qudits) $R_1,\ldots,R_M$ if all qudits $Q_1,\ldots,Q_N$ are in the unit state.
This gate can be implemented with the same scheme as the \CnZ~gate with a modification in the basic operation depicted in Fig.~\ref{fig:generalization}(c).
Here the central gate is~\CUb$^{(R_1,\ldots, R_M)}$ with a control on the tree root, and we apply triples $X_m$, \CX, $X$ on all the leaves $11,\ldots,1n(1)$.
We note that in this scheme the tree is constructed in the space of control qudits $Q_1,\ldots, Q_N$ and the dimension of the root qudit space has to be at least $2+n(1)$.
One can see that the whole scheme requires $2N-2+N_{\textsf{C}{\bf U}}$ two-qudit operations, where $N_{\textsf{C}{\bf U}}$ is the number of two-qudit operations required for performing  the \CUb$^{(R_1,\ldots,R_M)}$~gate. 

\section{Experimental realizations} \label{sec:experiment}

Qudit ensembles can be created and controlled in experiments with quantum systems of various nature.
The qudits systems have already been demonstrated in superconducting systems~\cite{Gustavsson2015,Katz2015,Ustinov2015}, integrated optics~\cite{White2009,Morandotti2017}, and NMR setups~\cite{Balestro2017}.
Other promising setups can be arrays of neutral atoms in optical tweezers and ions in linear traps.
In these systems, one can encode qudits in different Zeeman states of the ground hyperfine state~\cite{Saffman2014}.
In particular, for the case of one-dimensional atomic array one can use qutrits in the following sequence of states $(F = 2; m_F = -2), (F = 1; m_F = -1), (F = 2; m_F = 0), (F=1; m_F= 1), (F=2, m_F =2)$, 
where $F$ is the total angular momentum vector and $m_F$ is its $z$-axis projection.
The single qudit operations can be done with microwave pulses or Raman transitions, such as in the case of $^{137}{\rm Ba}^{+}$ ions, where for five-level qudits the estimated single-qudit fidelity is on the level of 99\%~\cite{Senko2019}.
Specifically, $^{137}{\rm Ba}^{+}$ ions have a long-lived state D$_{5/2}$, and do not require an octupole transition during qudit-state measurements.
The high fidelity two-qudits entangling gates can be realized with the qudit Molmer--Sorensen gate~\cite{Senko2019} in trapped ion systems and with Rydberg blockade in atomic arrays~\cite{Lukin20192,Saffman2019}.

\section{Conclusion and outlook} \label{sec:concl}

We have demonstrated that a strong reduction in the number of operations and in the depth of quantum circuits can be achieved by using qudit systems satisfying a certain relation between their dimensionality and topology. 
This is of importance for an efficient implementation of a generalized Toffoli gate as part of the algorithms.
A clear example is the diffusion operator in Grover's algorithm~\cite{Grover1996}, i.e., an operator acting after each appeal to an oracle.
It requires an $n$-qubit Toffoli gate, where $n$ is the length of input for the oracle. 
Another example is the employment of the generalized Toffoli gate in the recently proposed artificial neuron quantum circuit~\cite{Tacchino2019}.
There are also proposals for employing generalized Toffoli gates in an increment circuit, which can be used for efficient implementation of Shor's algorithm~\cite{Gokhale2019}.
With the reduced number of operations in the case of using qudits, one can expect a significant speed-up in the realization of these algorithms. 

Toffoli gates are also key ingredients for the realization of quantum error-correction codes~\cite{Shor1996,Cory1998,Reed2012}.
Qutrits are already being used for efficient realization of Toffoli gates in superconducting qubit systems~\cite{Reed2012}.
In this direction, our method paves a way for the reduction of the cost of the error-correction procedure and the implementation of more complicated codes~\cite{Zhang2012}.

\section*{Acknowledgments}

We are grateful to B.L. Altshuler, R. Blatt, and V.I. Yudson for valuable discussions.
The work was supported by the Russian Science Foundation Grant No. 19-71-10091 (Secs. II, III, and IV).
The research leading to these results has received funding from the European Research Council under European Community's Seventh Framework Programme (FP7/2007-2013 Grant Agreement No. 341197; experimental analysis in Sec. V).

\end{document}